\documentclass[12pt]{article}
\usepackage{epsfig,amsfonts,amssymb}

\newcommand{\be}{\begin{equation}}
\newcommand{\ee}{\end{equation}}
\newcommand{\bea}{\begin{eqnarray}}
\newcommand{\eea}{\end{eqnarray}}
\newcommand{\beas}{\begin{eqnarray*}}
\newcommand{\eeas}{\end{eqnarray*}}

\font\cmss=cmss12
\newcommand{\identity}{{\rlap{\cmss 1} \hskip 1.6pt \hbox{\cmss 1}}}

\newcommand{\phinotg} [1]{{\phi_{0#1}^{\it global}}}
\newcommand{\phinotgL}[1]{{\phi_{0#1}^{\it global,\,L}}}
\newcommand{\phinotgR}[1]{{\phi_{0#1}^{\it global,\,R}}}
\newcommand{\phinotp} [1]{{\phi_{0#1}^{\it Poincare}}}
\newcommand{\phinotr} [1]{{\phi_{0#1}^{\it Rindler}}}
\newcommand{\phinotrL}[1]{{\phi_{0#1}^{\it Rindler,\,L}}}
\newcommand{\phinotrR}[1]{{\phi_{0#1}^{\it Rindler,\,R}}}
\newcommand{\phinotk} [1]{{\phi_{0#1}^{\it Kruskal}}}

\begin{document}

\baselineskip 14 pt
\parskip 12 pt

\begin{titlepage}
\begin{flushright}
{\small BROWN-HET-1448} \\
{\small CU-TP-1130} \\
{\small hep-th/0506118}
\end{flushright}

\begin{center}

\vspace{2mm}

{\Large \bf Local bulk operators in AdS/CFT: \\
a boundary view of horizons and locality}

\vspace{3mm}

Alex Hamilton${}^1$, Daniel Kabat${}^1$, Gilad Lifschytz${}^2$ and David A.\ Lowe${}^3$

\vspace{2mm}

${}^1${\small \sl Department of Physics} \\
{\small \sl Columbia University, New York, NY 10027 USA} \\
{\small \tt hamilton, kabat@phys.columbia.edu}
\vspace{1mm}

${}^2${\small \sl Department of Mathematics and Physics and CCMSC} \\
{\small \sl University of Haifa at Oranim, Tivon 36006 ISRAEL} \\
{\small \tt giladl@research.haifa.ac.il}
\vspace{1mm}

${}^3${\small \sl Department of Physics} \\
{\small \sl Brown University, Providence, RI 02912 USA} \\
{\small \tt lowe@brown.edu}

\end{center}

\vskip 0.3 cm

\noindent
We develop the representation of local bulk fields in AdS by non-local
operators on the boundary, working in the semiclassical limit and
using AdS${}_2$ as our main example.  In global coordinates we show
that the boundary operator has support only at points which are
spacelike separated from the bulk point.  We construct boundary
operators that represent local bulk operators inserted behind the
horizon of the Poincar\'e patch and inside the Rindler horizon of a
two dimensional black hole.  We show that these operators respect bulk
locality and comment on the generalization of our construction to
higher dimensional AdS black holes.

\end{titlepage}

\section{Introduction}

The AdS/CFT correspondence provides a non-perturbative definition of
string theory in asymptotically AdS space \cite{Maldacena, AdSreview}.
In principle all bulk observables are encoded in correlation functions
of local operators in the CFT.  In practice, however, many of the
quantum gravity questions we would like to address are not simply related
to local boundary correlators.  These questions include: how does a
quasi-local bulk spacetime emerge from the CFT?  How does the region
behind a horizon get encoded in the CFT?  What is the CFT description
(or perhaps resolution) of a black hole singularity?

In this paper we develop a set of tools for recovering local bulk
physics from the CFT.  We use the Lorentzian AdS/CFT correspondence
developed in \cite{Balasubramanian:1998sn,BDHM,Marolf}.  The basic idea is to express local
operators in the bulk in terms of non-local operators on the boundary.
We work in the leading semiclassical approximation -- meaning both
large $N$ and large 't Hooft coupling -- and consider free scalar
fields in AdS.  The fields are taken to have normalizable fall-off
near the boundary of AdS.
\[
\phi(z,x) \sim z^\Delta \phi_0(x)~.
\]
Here $z$ is a radial coordinate which vanishes at the boundary.
General AdS/CFT considerations imply that the boundary behavior of the
field corresponds to an operator of conformal dimension $\Delta$ in
the CFT.
\[
\phi_0(x) \leftrightarrow {\cal O}(x)~.
\]
This implies a correspondence between local fields in the bulk and
{\em non-local} operators in the CFT.
\[
\phi(z,x) \leftrightarrow \int dx' \, K(x' \vert z,x) {\cal O}(x')~.
\]
We will refer to the kernel $K(x' \vert z,x)$ as a smearing function.
Bulk-to-bulk correlation functions, for example, are then equal to
correlation functions of the corresponding non-local operators in the
dual CFT
\[
\langle \phi(z_1,x_1) \phi(z_2,x_2) \rangle = \int dx_1' dx_2'
K(x_1' \vert z_1,x_1) K(x_2' \vert z_2,x_2) \langle {\cal O}(x_1')
{\cal O}(x_2') \rangle~.
\]
In this paper we construct smearing functions and show how certain
aspects of bulk physics are encoded by these non-local operators.  We
will use AdS${}_2$ as our main example, although many of the results
presented here generalize to higher dimensions \cite{ToAppear}.
Smearing functions were discussed in \cite{Balasubramanian:1999ri},
and smearing functions in AdS${}_5$, as well as in some non-conformal
variants, have been computed by Bena \cite{Bena}.  An algebraic
formulation of the correspondence between local bulk fields and
non-local boundary observables was developed in \cite{Rehren}.  Other
studies of bulk locality and causality include \cite{bulk}.

To avoid any possible confusion, we note that Witten \cite{Witten} (see also \cite{gkp})
introduced a bulk-to-boundary propagator whose Lorentzian continuation
can be used to represent bulk fields having a prescribed {\em
non-normalizable} behavior near the boundary
\cite{Balasubramanian:1998sn}.  Such bulk fields are dual to sources
that deform the CFT action.  Our approach is quite different: we work
directly with the undeformed CFT, and introduce non-local operators
that are dual to {\em normalizable} fluctuations in the bulk of
AdS.\footnote{In Witten's approach one can relate expectation values
of local operators in the CFT to the normalizable boundary behavior of
fields in AdS \cite{Freedman:1998tz,KlebanovWitten}.}

An outline of this paper is as follows.  In section 2 we construct
smearing functions in AdS${}_2$ in global coordinates.  In section 3
we construct smearing functions in Poincar\'e coordinates and show how
the Poincar\'e horizon appears in the CFT.  In section 4 we discuss
the way in which bulk-to-bulk correlators are recovered from the CFT
and show how coincident singularities arise.  In section 5 we work in
Rindler coordinates and discuss the appearance of black hole horizons.
In section 6 we extend the picture to general AdS black holes.  We
conclude in section 7.

\section{Global smearing functions}

\subsection{AdS${}_2$ generalities}

We begin by reviewing a few standard results; for more details see
\cite{Balasubramanian:1998sn} or appendix A.  In global coordinates
the AdS${}_2$ metric is
\be
ds^2 = {R^2 \over \cos^2 \rho} \left(- d\tau^2 + d\rho^2 \right)
\ee
where $R$ is the radius of curvature and $-\infty < \tau < \infty$,
$-\pi/2 \leq \rho < \pi/2$.  It is convenient to introduce a distance
function
\be
\label{DistanceFtn}
\sigma(\tau,\rho \vert \tau',\rho') = {\cos (\tau - \tau') - \sin \rho \sin \rho' \over \cos \rho \cos \rho'}
\ee
which is invariant under AdS isometries.  Points that can be connected
by a geodesic necessarily lie in the unit cell $-\pi < \tau - \tau' <
\pi$.  For such points
\[
\sigma = \left\lbrace
\begin{array}{ll}
\cos(s/R) & \quad\hbox{\rm timelike ($s$ = geodesic proper time)} \\
1 & \quad\hbox{\rm null} \\
\cosh(d/R) & \quad\hbox{\rm spacelike ($d$ = geodesic proper distance)~.}
\end{array}
\right.
\]
Points in the unit cell with $\sigma < -1$ are timelike separated but
are not connected by a geodesic.  A free scalar field of mass $m$ can
be expanded in a complete set of normalizable modes\footnote{For
certain masses there are inequivalent sets of normalizable modes, a
complication analyzed in \cite{BreitenlohnerFreedman, KlebanovWitten}
which we will not consider further here.}
\be
\label{BulkModeExpansion}
\phi(\tau,\rho) = \sum_{n=0}^\infty a_n e^{- i \omega_n \tau} \cos^\Delta \rho \, C_n^\Delta(\sin \rho) + {\rm h.c.}
\ee
where $\omega_n = n + \Delta$, $\Delta = {1 \over 2} + \sqrt{{1 \over
4} + m^2 R^2}$ is the conformal dimension of the corresponding
operator, and $C_n^\Delta(x)$ is a Gegenbauer polynomial.  We have not
bothered normalizing the modes.

The field vanishes at the boundary of AdS.  In global coordinates we
define the right boundary value of the field by
\be
\label{GlobalPhi0}
\phinotgR{}(\tau) = \lim_{\rho \rightarrow \pi/2} \,\, {\phi(\tau,\rho) \over \cos^\Delta \rho}\,.
\ee
Similarly the left boundary value is
\be
\phinotgL{}(\tau) = \lim_{\rho \rightarrow - \pi/2} \,\, {\phi(\tau,\rho) \over \cos^\Delta \rho}\,.
\ee
Some special simplifications occur when $\Delta$ is a positive
integer.  First of all, in this case the field is single-valued on the
AdS${}_2$ hyperboloid (meaning that we can identify $\tau \approx \tau
+ 2 \pi$).  Also we define the antipodal map on AdS${}_2$
\be
\label{antipodal}
A \, : \, (\tau,\rho) \mapsto (\tau + \pi, -\rho)\,.
\ee
Note that $\sigma(x \vert Ax') = - \sigma(x \vert x')$.  When $\Delta$
is a positive integer we have
\[
\phi(Ax) = (-1)^\Delta \phi(x)
\]
in which case the boundary values are related by
\be
\label{reflect}
\phinotgL{}(\tau) = (-1)^\Delta \phinotgR{}(\tau + \pi)\,.
\ee
%

\subsection{Green's function approach}
\label{GreensSect}

In this subsection we construct smearing functions for AdS${}_2$ in
global coordinates starting from a suitable Green's function.

The Green's function should satisfy
\be
\label{GreensEqn}
\left(\Box - m^2\right) G(x \vert x') = {1 \over \sqrt{-g}} \delta^2(x - x')\,,
\ee
where $\delta^2(x-x')$ is defined on the universal cover of AdS, $-\pi/2\leq \rho<\pi/2$, $-\infty<\tau<\infty$.
We want a smearing function that is non-zero only at spacelike
separation, so we make the ansatz
\[
G(x \vert x') = f\left(\sigma(x \vert x')\right) \theta\Bigl((\rho - \rho')^2 - (\tau - \tau')^2\Bigr)\,.
\]
Here $\sigma$ is the AdS invariant distance defined in
(\ref{DistanceFtn}).  Due to the step function $G$ is non-zero only at
spacelike separation.  By direct substitution one can check that
(\ref{GreensEqn}) is satisfied provided that $f(\sigma)$ satisfies the
homogeneous AdS-invariant wave equation
\be
(\sigma^2 - 1) f^{\prime\prime}(\sigma) + 2 \sigma f^\prime(\sigma) - \Delta (\Delta - 1) f(\sigma) = 0
\ee
with the boundary condition $f(1) = 1/4$.\footnote{To see this use the
AdS invariance to set $\tau' = \rho' = 0$, work in light-front
coordinates $x^\pm = \tau \pm \rho$, and write the step function as
$\theta(x^+) \theta(-x^-) + \theta(x^-) \theta(-x^+)$.}  The solution
\[
G(x \vert x') = {1 \over 4} P_{\Delta - 1}(\sigma) \, \theta\Bigl((\rho - \rho')^2 - (\tau - \tau')^2\Bigr)
\]
is given by a Legendre function.  It is worth emphasizing some curious
properties of this Green's function.  First of all, by construction
it is non-zero only at spacelike separation.  It is finite (but
discontinuous) on the light cone, with $G \rightarrow 1/4$ as the
light cone is approached from a spacelike direction.  However it is
non-normalizable near the boundary of AdS, with
\[
G(x \vert x') \sim {\Gamma(2 \Delta -1) \over 2^{\Delta + 1} \Gamma(\Delta)^2} \,
\sigma^{\Delta - 1}
\]
at large spacelike separation.  In AdS${}_2$ we can simplify the
discussion by working with a Green's function that is non-zero only in
the right-hand part of the light cone.\footnote{We could have chosen
the left part of the light cone.  In either case, this step is not
possible in higher dimensions.}  With similar arguments it is easy to
see that
\be
\label{Greens}
G(x \vert x') = {1 \over 2} P_{\Delta - 1}(\sigma) \theta(\rho - \rho') \theta(\rho - \rho' - \vert \tau - \tau' \vert)
\ee
is a suitable Green's function.

Having constructed a Green's function that is non-zero only in the right light cone
we can make use of Green's identity
\[
\phi(x') = \int_{-\infty}^\infty d\tau \sqrt{-g} \, \left[\phi(\tau,\rho) \partial^\rho G(\tau,\rho \vert x') -
G(\tau,\rho \vert x') \partial^\rho \phi(\tau,\rho)\right]
\vert_{\rho = \rho_0}\,.
\]
We are interested in sending the regulator $\rho_0 \rightarrow \pi/2$.
In this limit only the leading behavior of both the field and the
Green's function contributes, and we have
\be
\phi(x') = \int_{-\infty}^\infty d\tau K(\tau \vert x') \phinotgR{}(\tau)
\ee
where the smearing function can be variously expressed as
\bea
\label{GreensRep}
K(\tau \vert \tau',\rho') & = & (2\Delta - 1) \lim_{\rho \rightarrow \pi/2}
\cos^{\Delta - 1} \rho \, G(\tau,\rho \vert \tau',\rho') \\
\label{conformal}
& = & {2^{\Delta -1} \Gamma(\Delta + 1/2) \over \sqrt{\pi} \, \Gamma(\Delta)}
\lim_{\rho \rightarrow \pi/2} (\sigma \cos \rho)^{\Delta-1} \theta(\rho - \rho' - \vert \tau - \tau' \vert) \\
\nonumber
& = & {2^{\Delta -1} \Gamma(\Delta + 1/2) \over \sqrt{\pi} \, \Gamma(\Delta)}
\left({\cos(\tau - \tau') - \sin \rho' \over \cos \rho'}\right)^{\Delta-1}
\theta\left({\pi \over 2} - \rho' - \vert \tau - \tau' \vert \right)~. \\
\label{finite}
\eea
These smearing functions have several important properties.
\begin{itemize}
\item The smearing function has compact support on the boundary of
AdS: $K$ is non-zero on the boundary only within the right lightcone
of the point $(\tau',\rho')$.  Of course we could have chosen to
construct a smearing function that was non-zero only in the left
lightcone.  Note that a local bulk operator near the left boundary
could be described with a highly localized smearing function on the
left boundary, or with a delocalized smearing function on the right
boundary.
\item The whole set-up is AdS covariant, since $\phinotgR{}$
transforms as a primary field with dimension $\Delta$ under conformal
transformations.  This is clear when $K$ is written in the form
(\ref{conformal}): the factor $\cos ^{\Delta - 1}\rho$ appearing in
that expression cancels the conformal weight of the field together
with the conformal weight of the measure $\int d\tau$.
\item
As can be seen explicitly in (\ref{finite}), the smearing function has
a finite limit as the regulator is removed, $\rho_0 \rightarrow
\pi/2$.
\end{itemize}
Note that these properties all follow from the fact that we began with
a Green's function that is non-normalizable near the boundary and
non-zero only at spacelike separation.  We have plotted the $\Delta = 3$
smearing function in Fig.~1.  As a simple case, note that for a
massless field in AdS${}_2$ one has $\Delta = 1$ and the general
expression reduces to
\[
K = {1 \over 2} \, \theta\Bigl({\pi \over 2} - \rho' -
\vert \tau - \tau' \vert \Bigr)\,.
\]
That is, a massless bulk field is expressed in terms of its boundary
value by
\[
\phi(\tau',\rho') = {1 \over 2} \int_{\tau' - (\pi/2 - \rho')}^{\tau' + (\pi/2 - \rho')}
d\tau \, \phinotgR{}(\tau)\,.
\]

\begin{figure}
\centerline{\epsfig{file=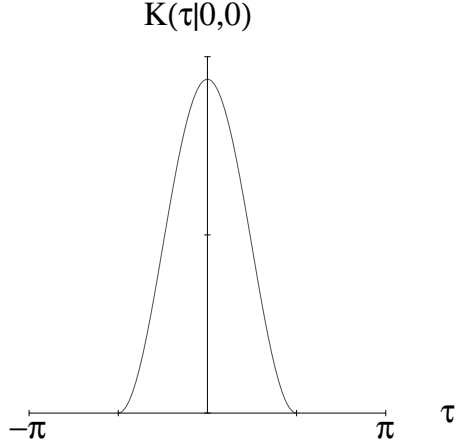}}
\caption{Global smearing function for a $\Delta = 3$ bulk operator located at
$\rho = 0$.}
\end{figure}

\subsection{Mode sum approach}

In this subsection we take a different point of view and construct
global smearing functions from a mode sum.  We begin with integer
conformal dimension then generalize.

Let us first suppose that $\Delta$ is a positive integer.  Then given
an on-shell bulk field with mode expansion (\ref{BulkModeExpansion})
we can reconstruct the bulk field from its right boundary value using
\be
\label{an}
a_n = {1 \over C_n^\Delta(1)} \oint {d\tau \over 2\pi} \, e^{i \omega_n \tau} \phinotgR{}(\tau)\,.
\ee
The integral is over any $2\pi$ interval on the boundary.  Plugging
this back into the bulk mode expansion, we can write (as an operator
identity!)
\be
\label{OperatorId}
\phi(\tau',\rho') = \oint d\tau \, K(\tau \vert \tau',\rho')
\phinotgR{}(\tau)
\ee
where the smearing function
\be
\label{AdS2sum}
K(\tau \vert \tau',\rho') = {1 \over 2 \pi} e^{i \Delta (\tau - \tau')} \cos^\Delta \rho'
\sum_{n=0}^\infty e^{i n (\tau - \tau' + i \epsilon)}
{C_n^\Delta(\sin \rho') \over C_n^\Delta(1)} + {\rm c.c.}
\ee
We have inserted an $i \epsilon$ to keep the mode sum convergent.  Note
that $K$ is periodic in $\tau$, with the same periodicity as the
underlying modes, unlike the Green's function (\ref{Greens}).  To make
contact with the results of the previous section we will eventually
choose the range of integration in (\ref{OperatorId}) to be $-\pi < \tau -
\tau' < \pi$.

It is important to note that the smearing functions are not unique.
For example we could equally well have constructed a smearing function
on the left boundary.  More importantly, from
(\ref{BulkModeExpansion}) note that $\phinotgR{}$ does not have Fourier
components with frequencies in the range $-\Delta+1,\, \cdots ,\,
\Delta - 1$, so we are free to drop any Fourier components of $K$ with
frequencies in this range.

To evaluate $K$ we use the integral representation \cite{GandR}
\[
{C_n^\Delta(x) \over C_n^\Delta(1)} = {\Gamma(\Delta + 1/2) \over \sqrt{\pi} \, \Gamma(\Delta)}
\int_0^\pi d\theta \, \sin^{2 \Delta - 1} \theta \,\,
\bigl(x + \sqrt{x^2 - 1} \cos \theta\bigr)^n~.
\]
Performing the sum on $n$ gives
\[
K(\tau \vert 0,\rho') = {1 \over 2 \pi} e^{i \Delta \tau} \cos^\Delta \rho'
{\Gamma(\Delta + 1/2) \over \sqrt{\pi} \, \Gamma(\Delta)}
\int_0^\pi d\theta \, {\sin^{2 \Delta - 1} \theta \over 1 - e^{i(\tau + i \epsilon)}
(\sin \rho' + i \cos \rho' \cos \theta)} + {\rm c.c.~.}
\]
The integral is a polynomial in $e^{i \tau}$ plus a logarithm
\cite{GandR}.  The polynomial only involves Fourier modes which do not
appear in $\phinotg{}$, so we may drop it leaving
\beas
K(\tau \vert 0,\rho') & \simeq & {2^{\Delta -1} \Gamma(\Delta + 1/2) \over \sqrt{\pi} \, \Gamma(\Delta)}
\left({\cos \tau - \sin \rho' \over \cos \rho'}\right)^{\Delta - 1} \\
&& \qquad \times \, {i \over 2 \pi} \, \log \left(
{1 - i e^{i(\tau - \rho' + i \epsilon)} \over 1 + i e^{-i (\tau - \rho' - i \epsilon)}}\,
{1 - i e^{-i(\tau + \rho' - i \epsilon)} \over 1 + i e^{i (\tau + \rho' + i \epsilon)}}\right)~.
\eeas
Here $\simeq$ means up to terms whose time dependence is such that
they vanish when integrated against $\phinotg{}$.  At this point it
helps to note that $f(x) = -i\log {1 + e^{i(x + i \epsilon)} \over 1 +
e^{-i(x - i \epsilon)}}$ is a sawtooth function, $f(x) = x$ for $-\pi
< x < \pi$ and $f(x + 2 \pi) = f(x)$.  Again some Fourier modes do not
contribute -- only the discontinuities of the sawtooth function matter
-- and one is left with
\be
\label{AdS2smear}
K(\tau \vert \tau',\rho') \simeq {2^{\Delta -1} \Gamma(\Delta + 1/2) \over \sqrt{\pi} \, \Gamma(\Delta)}
\left({\cos (\tau - \tau') - \sin \rho' \over \cos \rho'} \right)^{\Delta - 1}
\theta\Bigl(\cos(\tau - \tau') - \cos\big({\pi \over 2} - \rho'\big)\Bigr)~.
\ee
Choosing the range of integration in (\ref{OperatorId}) to be $-\pi <
\tau - \tau' < \pi$, the step function in (\ref{AdS2smear}) reduces to
the step function in (\ref{finite}), so this expression reproduces the
result we obtained in the previous subsection starting from a Green's
function.\footnote{The smearing function defined in (\ref{AdS2sum})
does not transform nicely under AdS isometries.  One can see this, for
example, by doing the sum explicitly for a massless field.
Fortunately the non-uniqueness of the smearing function allowed us to
construct an AdS covariant expression.}

What if $\Delta$ is not an integer?  Recall we are working on the universal cover of AdS where $-\infty<\tau<\infty$. In this case the mode functions
are no longer periodic.  This means (\ref{an}) is no longer valid,
since for general $\Delta$ the positive and negative frequency modes
are not orthogonal on the interval $-\pi < \tau < \pi$.  The trick is
to first decompose the field into positive and negative frequency
pieces, $\phi(\tau,\rho) = \phi_+(\tau,\rho) + \phi_-(\tau,\rho)$
where
\be
\label{PosNeg}
\phi_+(\tau,\rho) = \sum_{n = 0}^\infty a_n e^{-i \omega_n \tau} \cos^\Delta \rho \,
C_n^\Delta(\sin \rho)
\ee
and $\phi_- = \phi_+^*$.  We can recover $\phi_\pm$ from their
boundary values, integrating over only the range
$-\pi\leq\tau-\tau'<\pi$, via $\phi_\pm(\tau',\rho') = \oint d\tau
K_\pm(\tau \vert \tau',\rho') \phinotgR{\pm}(\tau)$ where
\be
\label{Kpm1}
K_\pm(\tau \vert \tau',\rho') = {1 \over 2\pi} \cos^\Delta \rho' \sum_{n = 0}^\infty
e^{\pm i \omega_n (\tau - \tau' \pm i \epsilon)} {C_n^\Delta(\sin \rho') \over C_n^\Delta(1)}\,.
\ee
These positive and negative frequency smearing functions are highly
non-unique, since for example we can add Fourier modes to $K_+$ that
$\sim e^{i (\Delta - 1)(\tau - \tau')}$, $e^{i (\Delta - 2)(\tau -
\tau')}$, $\cdots$.  By making use of this freedom we can put $K_\pm$
into the form of an image sum,
\be
\label{Kpm2}
K_\pm \simeq \sum_{k = -\infty}^\infty e^{\pm i 2 \pi k \Delta} K(\tau \vert \tau' + 2 \pi k,\rho')
\ee
where for any $\Delta$
\be
K(\tau \vert \tau',\rho') \simeq {2^{\Delta -1} \Gamma(\Delta + 1/2) \over \sqrt{\pi} \, \Gamma(\Delta)}
\left({\cos (\tau - \tau') - \sin \rho' \over \cos \rho'} \right)^{\Delta - 1}
\theta\Bigl({\pi \over 2} - \rho' - \vert \tau - \tau' \vert \Bigr)
\ee
is defined over the range $-\infty<\tau-\tau'<\infty$.  One can verify (\ref{Kpm2}) by doing an
inverse Fourier transform using a complete set of skew-periodic modes
$e^{\pm i (n + \Delta) (\tau - \tau')}$, $n \in {\mathbb Z}$, and
showing that the coefficients of the $n \geq 0$ modes appear in
(\ref{Kpm1}).  On the restricted interval $-\pi < \tau - \tau' < \pi$ only the $k=0$ term contributes, so the
smearing functions all agree and
\[
\phi = \int_{\tau' - \pi}^{\tau' + \pi} d\tau \, \left(K_+ \phinotgR{+} + K_- \phinotgR{-}\right) =
\int_{\tau' - \pi}^{\tau' + \pi} d\tau \, K \phinotgR{}
\]
reproduces our previous Green's function expression.

\section{Poincar\'e coordinates}

\subsection{Smearing functions}

It is worth asking how these non-local operators look in
different coordinate systems.  For example one can introduce
Poincar\'e coordinates
\bea
&&Z = {R \cos \rho \over \cos \tau + \sin \rho} \qquad
T = {R \sin \tau \over \cos \tau + \sin \rho} \\[10pt]
\nonumber
&& \quad 0 < Z < \infty \qquad\qquad -\infty < T < \infty
\eea
in which
\[
ds^2 = {R^2 \over Z^2} \left(-dT^2 + dZ^2\right)\,.
\]
These coordinates only cover an interval $-\pi < \tau < \pi$ of global
time on the right boundary.  The mode expansion reads
\[
\phi(T,Z) = \int_0^\infty d\omega \, a_\omega e^{-i \omega T} \sqrt{Z}
J_\nu(\omega Z) + {\rm c.c.}
\]
where $\nu = \Delta - {1 \over 2}$.  The field vanishes as $Z
\rightarrow 0$.  In Poincar\'e coordinates it is convenient to define
the boundary value
\be
\label{PoincareBdy}
\phinotp{}(T) = \lim_{Z \rightarrow 0} {\phi(T,Z) \over Z^\Delta} \,.
\ee
Note that this differs from the definition (\ref{GlobalPhi0}) we used
in global coordinates.

It is straightforward to compute the smearing function in Poincar\'e
coordinates directly from a mode sum.  Given an on-shell bulk field we
have
\[
a_\omega = \int_{-\infty}^\infty dT \, e^{i \omega T} \, {1 \over 2 \pi
\sqrt{Z} J_\nu(\omega Z)} \phi(T,Z)~,
\]
or taking the limit $Z \rightarrow 0$
\[
a_\omega = {\Gamma(\nu + 1) \over 2 \pi (\omega/2)^\nu} \int_{-\infty}^\infty dT \,
e^{i \omega T} \phinotp{}(T)\,.
\]
Plugging this back into the bulk mode expansion we can write
\[
\phi(T',Z') = \int_{-\infty}^\infty dT \, K(T \vert T',Z') \phinotp{}(T)
\]
where the Poincar\'e smearing function is
\be
\label{Poincare}
K(T \vert T',Z') = {1 \over \pi} 2^\nu \Gamma(\nu + 1) \sqrt{Z'}
\int_0^\infty d\omega \, {1 \over \omega^\nu} J_\nu(\omega Z') \cos \omega(T - T')\,.
\ee
The integral is well defined without an $i \epsilon$ prescription.  It
vanishes identically for $\vert T - T' \vert > Z'$, while for $\vert T
- T' \vert < Z'$ we have \cite{GandR}
\[
K = {\Gamma(\Delta + 1/2) \over \sqrt{\pi} \Gamma(\Delta)} \, Z'^{\Delta - 1}
F\left({1 \over 2},1 - \Delta, {1 \over 2}, {(T - T')^2 \over Z'{}^2}\right)\,.
\]
Thus in general
\be
K = {\Gamma(\Delta + 1/2) \over \sqrt{\pi} \, \Gamma(\Delta)}
\left({Z'{}^2 - (T - T')^2 \over Z'}\right)^{\Delta-1}
\theta(Z' - \vert T - T' \vert)\,.
\ee
where we used $F(\alpha,\beta,\alpha,x) = (1 - x)^{-\beta}$.  This
expression is valid for any positive $\Delta$.  Note that, unlike the
global mode sum (\ref{AdS2sum}), the Poincar\'e mode sum is non-zero
only at spacelike separation.  It is also AdS covariant since
\be
\label{PoincareCov}
K(T \vert T',Z') = {2^{\Delta -1} \Gamma(\Delta + 1/2) \over \sqrt{\pi} \, \Gamma(\Delta)}
\lim_{Z \rightarrow 0} \bigl(Z \sigma(T,Z \vert T',Z')\bigr)^{\Delta-1} \theta(\sigma-1)
\ee
where the AdS invariant distance (\ref{DistanceFtn}) expressed in
Poincar\'e coordinates is
\be
\label{DistanceFtn2}
\sigma(Z,T \vert Z',T') = {Z^2 + Z'{}^2 - (T - T')^2 \over 2ZZ'}\,.
\ee
Upon changing variables from Poincar\'e time to global time, this is
equivalent to our previous result (\ref{conformal}).  To see this, one
merely has to note that near the boundary
\[
Z^\Delta \phinotp{} \sim \cos^\Delta \rho \, \phinotgR{}
\]
while the change of integration measure is
\[
{dT \over Z} = {d\tau \over \cos \rho}\,.
\]

\subsection{Going behind the Poincar\'e horizon}\label{behindp} 

In the previous subsection we showed how local bulk fields in the
Poincar\'e patch are represented by smeared operators on the boundary.
But what if the bulk point is outside the Poincar\'e patch?  Can it
still be represented as an operator on the boundary of the Poincar\'e
patch?  The answer turns out to be yes: we can still work in
Poincar\'e coordinates on the boundary, but we have to use a different
smearing function.

To obtain the correct smearing function our strategy is to start with
the global smearing function, manipulate it so that it is non-zero only
on the boundary of the Poincar\'e patch, and then convert to
Poincar\'e coordinates.  We first assume that $\Delta$ is a positive
integer.  In this case $\phinotgR{}(\tau)$ is periodic in the global
time coordinate $\tau$ with period $2\pi$, so one can take the global
smearing function (\ref{AdS2smear}) and translate whatever part of it
has left the Poincar\'e patch by a multiple of $2\pi$ in order to get
it back inside the Poincar\'e patch.  This is illustrated in Fig.~2.

\begin{figure}
\centerline{\epsfig{file=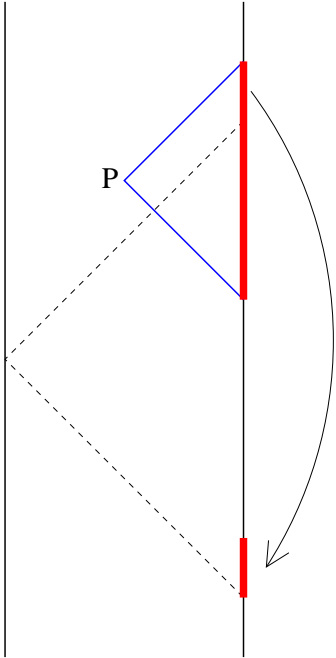,height=250pt}}
\caption{Smearing function for a bulk operator with integer $\Delta$
located outside the Poincar\'e patch.}
\end{figure}

This can be expressed quite simply in terms of the invariant distance
(\ref{DistanceFtn}).  Noting that $\sigma$ is $2\pi$ periodic in
global time, for a general point $P$ (not necessarily inside the
Poincar\'e patch) we can express the smearing function in a form that,
upon changing to Poincar\'e coordinates, looks identical to
(\ref{PoincareCov}):
\bea
\label{Poincare2}
&&\phi(P) = \int_{-\infty}^\infty dT \, K(T \vert P) \phinotp{}(T) \\
\nonumber
&&K(T \vert P) = {2^{\Delta -1} \Gamma(\Delta + 1/2) \over \sqrt{\pi} \, \Gamma(\Delta)}
\lim_{Z \rightarrow 0} \bigl(Z\sigma(T,Z \vert P)\bigr)^{\Delta-1} \theta(\sigma-1)~.
\eea
Note that the signal of a bulk operator approaching the future (past)
Poincar\'e horizon is that the smearing function extends to Poincar\'e
time $T \rightarrow + \infty$ ($T \rightarrow - \infty$).

What happens if $\Delta$ is not an integer?  The trick is to note
that, although the field itself is not periodic in $\tau$, its
positive and negative frequency components (\ref{PosNeg}) are periodic
up to a phase:
\[
\phi_\pm(\tau + 2 \pi, \rho) = e^{\mp i 2 \pi \Delta} \phi_\pm(\tau,\rho)\,.
\]
For a bulk point $P$ with global coordinates $(\tau',\rho')$ we set
$\tau' = \tau'_0 + 2 \pi n$ with $-\pi < \tau'_0 < \pi$ and $n \in
{\mathbb Z}$ and write
\be
\label{PoinGlobNonZ}
\phi(\tau',\rho') = \int d\tau \, K_{\it global}(\tau \vert \tau_0',\rho')
\left(e^{-i 2 \pi n \Delta} \phinotgR{+}(\tau) + e^{i 2 \pi n \Delta}
\phinotgR{-}(\tau)\right)\,.
\ee
Converting to Poincar\'e coordinates this becomes
\be
\label{PoinPoinNonZ}
\phi(P) = \int_{-\infty}^\infty dT \, K_{\it Poincare}(T \vert P_0)
\left(e^{-i 2 \pi n \Delta} \phinotp{+}(T) + e^{i 2 \pi n \Delta}
\phinotp{-}(T)\right)\,.
\ee
Here $K_{\it Poincare}$ is given in (\ref{Poincare2}) and $P_0$ is the
bulk point with global coordinates $(\tau_0',\rho')$.  In deriving
this we used the fact that the global and Poincar\'e vacua are
identical \cite{Danielsson:1998wt,SpradlinStrominger}.  This
equivalence is discussed in more detail in appendix B.

When expressed in terms of $\phinotp{+}$ and $\phinotp{-}$ the
smearing function consists of one or two disconnected blobs on the
boundary.\footnote{One blob if the bulk point is inside the Poincar\'e
patch or any of its images under $\tau \rightarrow \tau + 2 \pi$, two
blobs otherwise.}  However expressing $\phinotp{\pm}$ in terms of
$\phinotp{}$ is completely non-local.  For example the positive
frequency part of the Poincar\'e boundary field is
\bea
&&\phinotp{+}(T') = \int dT \, P_+(T \vert T') \phinotp{}(T) \\
\nonumber
&&P_+(T \vert T') = \int_0^\infty {d\omega \over 2\pi} \,
e^{i \omega (T - T' + i \epsilon)} = {i \over 2 \pi (T - T' + i \epsilon)} 
\eea
For a point outside the Poincar\'e patch the smearing function in
terms of $\phinotp{}$ is non-zero everywhere.  Just as for integer
$\Delta$, the signal of a bulk operator approaching the future (past)
Poincar\'e horizon is that the smearing function extends to Poincar\'e
time $T \rightarrow + \infty$ ($T \rightarrow - \infty$).

\section{Recovering bulk correlators}

In the previous section we defined a set of smearing functions which
enable us to reconstruct a normalizable bulk field from its behavior
near the boundary of AdS.  We obtained these smearing functions by
solving a wave equation in AdS space, so our expressions are valid for
any state of the field provided we are in the limit of semiclassical
supergravity where back-reaction of the field on the geometry can be
neglected.

Assuming the existence of a dual CFT, we can identify local operators
in the bulk supergravity with non-local operators in the CFT.
\[
\phi(\tau',\rho') \leftrightarrow \int d\tau \, K(\tau \vert \tau',\rho') {\cal O}(\tau)~.
\]
This correspondence should hold for any state of the field
(equivalently, any state of the CFT) provided we are in the limit of
semiclassical supergravity where back-reaction is negligible; note
that in this limit the smearing functions are independent of the
state.  We can use the correspondence to recover bulk supergravity
correlation functions from the CFT, for example
\be
\label{corresp}
{}_{S}\langle \psi \vert \phi(\tau,\rho) \phi(\tau^\prime,\rho^\prime) \vert \psi \rangle_{S} = \int ds \int ds^\prime
K(s \vert \tau,\rho) K(s^\prime \vert \tau^\prime, \rho^\prime) {}_{C}\langle \psi \vert {\cal O}(s) {\cal O}(s') \vert \psi \rangle_{C}\,.
\ee
Here $\vert \psi \rangle_S$ is any supergravity state and $\vert \psi
\rangle_C$ is the corresponding CFT state.  Although true by
construction,\footnote{This is perhaps easiest to see if one
represents the smearing functions and correlators using mode sums.} at
first sight this is a rather surprising identity.  For example in the
global AdS${}_2$ vacuum the bulk and boundary Wightman functions are
\cite{SpradlinStrominger}
\bea
\nonumber
&&\langle \phi(\tau,\rho) \phi(\tau',\rho') \rangle =
{\Gamma(\Delta) \over 2^{\Delta + 1} \sqrt{\pi} \Gamma(\Delta + 1/2)}
\sigma^{-\Delta} F\left({\Delta \over2},{\Delta + 1 \over 2};{2\Delta + 1 \over 2};
{1 \over \sigma^2}\right) \\
&&\langle {\cal O}(s) {\cal O}(s') \rangle =
{(-1)^\Delta \Gamma(\Delta) \over 2^{2\Delta+1} \sqrt{\pi} \,
\Gamma(\Delta + {1 \over 2}) \sin^{2\Delta} \left({s - s' \over 2}\right)}\,.
\eea
When $\sigma < -1$ (bulk points not connected by a geodesic) both the
left and right hand sides of (\ref{corresp}) are unambiguous.  To
continue into the regime $\sigma > -1$ one has to check that the $i
\epsilon$ prescriptions go through correctly.  The bulk Wightman
function is defined by a $\tau \rightarrow \tau - i \epsilon$
prescription, and fortunately by translation invariance of $K(s \vert
\tau,\rho)$ this is equivalent to the correct $s \rightarrow s - i
\epsilon$ prescription for the boundary Wightman function.

Given (\ref{corresp}) we are guaranteed that -- in the strict
semiclassical limit -- two non-local boundary operators will commute whenever
the bulk points are spacelike separated.  Bena \cite{Bena} checked
explicitly that the commutator vanishes from the boundary point of
view.

\subsection{Light-cone singularities in the bulk}

Correlation functions of these non-local boundary operators diverge whenever
the corresponding bulk points are coincident or light-like separated.
In this section we explain how these singularities arise from the
boundary point of view.

It may seem a little surprising that the correlators diverge at all,
since in field theory one usually introduces smeared operators to
avoid singular correlators.  To see what is going on let us look at how
these operators are constructed.  It is simplest to work in terms of a
mode sum in frequency space.  From (\ref{AdS2sum}) or (\ref{Poincare})
one sees that the smearing is done by multiplying each mode of the
boundary operator by a frequency dependent phase.  These phases act as
a regulator which makes the correlator with most other operators
non-singular.  However for a given smeared operator one can find
certain other smeared operators for which the phases cancel.  For two
such operators the correlator has a UV divergence.

To make this explicit consider the correlator of two bulk operators in
the Poincar\'e patch.  Working in frequency space on the boundary the
correlator is
\be
\label{Frequency}
\langle \phi(T,Z) \phi(T',Z') \rangle = \int_{-\infty}^\infty d\omega \,
K(\omega \vert T,Z) K(-\omega \vert T',Z') \langle {\cal O}(\omega) {\cal O}(-\omega) \rangle\,.
\ee
The Poincar\'e smearing function in frequency space can be read off
from (\ref{Poincare}),
\beas
K(\omega \vert T,Z) & = & 2^\nu \Gamma(\nu + 1) {\sqrt{Z} \over \vert \omega \vert^\nu}
J_\nu(\vert \omega \vert Z) e^{-i \omega T} \\
& \sim & {1 \over \vert \omega \vert^{\nu + 1/2}} \cos(\vert \omega \vert Z + {\rm const.}) e^{-i \omega T}
\eeas
at large $\vert \omega \vert$.  For an operator of dimension $\Delta$
the boundary correlator in frequency space behaves as $\langle {\cal
O}(\omega) {\cal O}(-\omega) \rangle \sim \vert \omega \vert^{2\nu}$.
For generic bulk points the integrand in (\ref{Frequency}) has
oscillating phases which regulate the behavior at large $\omega$.  But
whenever $(T,Z)$ and $(T',Z')$ either coincide or are light-like
separated the phases cancel and the integral diverges logarithmically
in the UV.

It is important to note that, even in the center of AdS, the lightcone
singularities of the bulk theory arise from the UV behavior of the
boundary theory.  This means, for example, that any attempt to put a
UV cutoff on the boundary theory will modify locality everywhere in
the bulk.

\section{AdS${}_2$ black holes}

\subsection{Smearing functions}
\label{BHsmear}

One can also introduce Rindler coordinates on AdS${}_2$
\bea
\nonumber
&& {r \over R \sqrt{M}} = {\cos \tau \over \cos \rho} \qquad
\tanh {t \sqrt{M} \over R} = {\sin \tau \over \sin \rho} \\[5pt]
\label{adsbh}
&& R \sqrt{M} < r < \infty \qquad -\infty < t < \infty \\[5pt]
\nonumber
&& ds^2 = - \left({r^2 \over R^2} - M\right) dt^2 +
\left({r^2 \over R^2} - M\right)^{-1} dr^2~.
\eea
The metric looks like a black hole,\footnote{In fact it is the
dimensional reduction of a BTZ black hole \cite{Achucarro:1993fd}.}
but the ``mass'' $M$ is just an arbitrary dimensionless parameter
which enters through the change of coordinates.  These coordinates
only cover an interval $-\pi/2 < \tau < \pi/2$ of global time on the
right boundary, but they can be extended in the usual way to cover an
identical interval on the left boundary.

What do smearing functions look like in Rindler coordinates?  This
depends on whether the bulk point is inside or outside the Rindler
horizon.  For a bulk point in the right Rindler wedge there is no
difficulty: we merely have to transform the global smearing function
function (\ref{GreensRep}) to Rindler coordinates.  This gives a
Rindler smearing function which is non-zero only at spacelike
separation on the right boundary.  Likewise for a bulk point in the
left Rindler wedge we can construct a smearing function that is
non-zero only at spacelike separation on the left boundary.  But what
if the bulk point is inside the horizon?  Then we need to modify the
smearing function.  Just as in Poincar\'e coordinates, our strategy
will be to start with the global smearing function, manipulate it so
that it is non-zero only on the boundary of the Rindler patch, and then
convert to Rindler coordinates.

The analysis is simplest when $\Delta$ is an integer.  Then we can use
(\ref{reflect}) to take those parts of the global smearing function
that have left the boundary of the Rindler patch and bring them back
inside.  However one obtains a smearing function with support on both
the left and right boundaries of AdS.  This is illustrated in Fig.~3.

\begin{figure}
\centerline{\epsfig{file=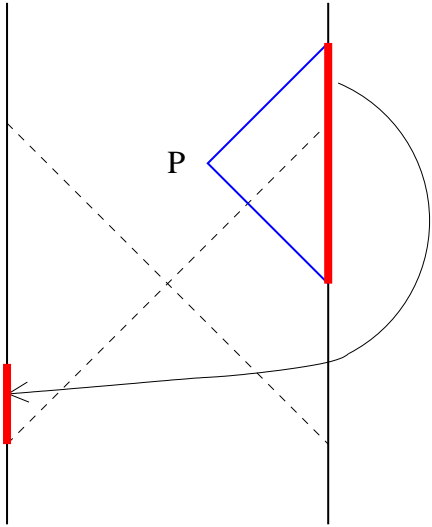,height=200pt}}
\caption{Smearing function for an integer-dimension bulk operator
located behind the Rindler horizon.}
\end{figure}

This can be expressed quite simply in terms of the invariant distance
$\sigma$.  Recall that $\sigma$ is $2\pi$ periodic in global time,
while under the antipodal map
\[
\sigma(x \vert Ax') = - \sigma(x \vert x') \qquad \phi(Ax) = (-1)^\Delta \phi(x)\,.
\]
By starting with the global smearing function (\ref{GreensRep}),
decomposing it into pieces that are inside / outside the Rindler
patch, and moving the outside part to the other boundary we have
\bea
\nonumber
&&\phi(P) = \int_{-\pi/2}^{\pi/2} d\tau \, \left(
K_{\it global}^R(\tau \vert P) \phinotgR{}(\tau) + (-1)^\Delta K_{\it global}^L(\tau \vert P) \phinotgL{}(\tau)
\right) \\
\label{GlobalRind}
&& K_{\it global}^R(\tau \vert P) = {2^{\Delta -1} \Gamma(\Delta + 1/2) \over \sqrt{\pi} \, \Gamma(\Delta)}
\lim_{\rho \rightarrow \pi/2} \bigl(\sigma(\tau,\rho \vert P) \cos \rho\bigr)^{\Delta-1} \theta(\sigma-1) \\
\nonumber
&& K_{\it global}^L(\tau \vert P) = {2^{\Delta -1} \Gamma(\Delta + 1/2) \over \sqrt{\pi} \, \Gamma(\Delta)}
\lim_{\rho \rightarrow -\pi/2} \bigl(-\sigma(\tau,\rho \vert P) \cos \rho\bigr)^{\Delta-1} \theta(-\sigma-1)~.
\eea
Here $P$ is any point inside the Rindler horizon.\footnote{In fact
this expression is valid for any $P$.  However if $P$ is in the left
Rindler wedge it is simpler to work with a smearing function that is
non-zero only at spacelike separation on the left boundary.}  To
express this in Rindler coordinates it is convenient to define the left
and right Rindler boundary fields $\phi_0^{\it Rindler,\,L/R}(t) =
\lim_{r \rightarrow \infty} r^\Delta \phi(t,r)$ where ${\it L}$ (${\it
R}$) refers to the left (right) Rindler wedge.  Near the boundary
\[
{\phi_0^{\it Rindler,\,L/R} \over r^\Delta} \sim \cos^\Delta \rho \, \phi_0^{\it global,\,L/R}
\]
while the change of integration measure is
\[
{d\tau \over \cos \rho} = {r dt \over R^2}\,.
\]
Putting this all together we have the final expression in Rindler coordinates
\bea
\nonumber
&&\phi(P) = \int_{-\infty}^{\infty} dt \, \left(K_{\it Rindler}^R(t \vert P) \phinotrR{}(t) +
(-1)^\Delta K_{\it Rindler}^L(t \vert P) \phinotrL{}(t) \right)\\
&& K_{\it Rindler}^R(t \vert P) = {2^{\Delta -1} \Gamma(\Delta + 1/2) \over \sqrt{\pi} \, \Gamma(\Delta) R^2}
\lim_{r \rightarrow \infty} (\sigma/r)^{\Delta-1} \theta(\sigma-1) \\
\nonumber
&& K_{\it Rindler}^L(t \vert P) = {2^{\Delta -1} \Gamma(\Delta + 1/2) \over \sqrt{\pi} \, \Gamma(\Delta) R^2}
\lim_{r \rightarrow \infty} (-\sigma/r)^{\Delta-1} \theta(-\sigma - 1)~.
\eea
Here $\sigma$ is the invariant distance from $P$ to the point $(t,r)$
on the appropriate boundary,
\[
\sigma(t,r\vert t',r') = {1 \over M} \left[{r r' \over R^2} \pm \left({r^2 \over R^2} - M\right)^{1/2}
\left(M - {r'{}^2 \over R^2}\right)^{1/2} \sinh {\sqrt{M}(t - t') \over R} \right]
\]
where $P$ is inside the horizon with coordinates $(t',r')$ and the
upper (lower) sign applies for $(t,r)$ near the right (left) boundary.
Note that the smearing function on the left boundary has support only
on points that are not connected to $P$ by a geodesic (points with
$\sigma < -1$).

What happens if $\Delta$ is not an integer?  Although we no longer
have (\ref{reflect}), a similar property holds separately for the
positive and negative frequency components of the boundary field
(\ref{PosNeg}):
\be
\phinotgR{\pm}(\tau) = e^{\pm i \pi \Delta} \phinotgL{\pm}(\tau + \pi)\,.
\ee
We can use this to rewrite the global smearing function in a form
similar to (\ref{GlobalRind}), where it is non-zero only on the
boundary of the Rindler patch.  For a bulk point $P$ inside the
horizon with global coordinates $(\tau',\rho')$ we set $\tau' =
\tau'_0 + 2 \pi n$, $n \in {\mathbb Z}$ and
\bea
\nonumber
&&\phi(P) = \int_{-\pi/2}^{\pi/2} d\tau \, \biggl[
  K_{\it global}^R(\tau \vert P) \Bigl(e^{-i 2 \pi n \Delta} \phinotgR{+}(\tau)
+ e^{i 2 \pi n \Delta} \phinotgR{-}(\tau)\Bigr) \\
\label{GeneralRind}
&& + K_{\it global}^L(\tau \vert P) \Bigl(e^{-i \pi (2n+1) \Delta} \phinotgL{+}(\tau)
+ e^{i \pi (2n+1) \Delta} \phinotgL{-}(\tau)\Bigr)
\biggr]
\eea
where $K_{\it global}^{R,L}(\tau \vert P)$ appear in
(\ref{GlobalRind}).  Expressed in terms of positive and negative
frequency global fields this smearing function consists of two
disconnected blobs, one on the left boundary and one on the
right.\footnote{For a bulk point in the right Rindler wedge or any of
its images under $\tau \rightarrow \tau + 2\pi$ one gets a single blob
on the right boundary.  For a bulk point in the left wedge or any of
its images it is more natural to use a smearing function with a single
blob on the left boundary.}  One can switch to Rindler coordinates,
however for points inside the horizon in general one obtains a
smearing function that is completely non-local in terms of $\phi_0^{\it
Rindler,\,L/R}$.  This is worked out in appendix B.

\subsection{Thermofield interpretation and black holes}

We can regard the AdS${}_2$ metric in Rindler coordinates as a
prototype for a black hole.  A standard Euclidean calculation shows
that the Hawking temperature is $1/\beta = {\sqrt{M} \over 2\pi R}$.
By keeping modes in both the left and right Rindler wedges, the
Hartle-Hawking state can be understood as a thermofield double
\cite{Israel:1976ur}.\footnote{Despite appearances, the Hartle-Hawking
state is independent of $M$.  In fact the global vacuum, the
Poincar\'e vacuum and the Hartle-Hawking state are identical
\cite{Lifschytz:1993eb,Danielsson:1998wt,SpradlinStrominger}.}

How does this look on the boundary?  As discussed in
\cite{Maldacena:2001kr} an eternal AdS black hole is dual to two
copies of the CFT in an entangled thermofield state.  As we show in
appendix A the thermofield Hamiltonian is
\[
H_{TF} = {\sqrt{M} \over R} (\hat{S}_1 \otimes \identity_2 - \identity_1 \otimes \hat{S}_2)\,.
\]
Here $\hat{S}$ is a non-compact conformal generator on the boundary.
The thermofield state is annihilated by $H_{TF}$, and can be formally
expressed in terms of $\hat{S}$ eigenstates as
\[
\vert \psi \rangle = {1 \over \sqrt{Z(\beta)}} \sum_n e^{-\beta E_n / 2} \vert n \rangle_1
\otimes \vert n \rangle_2\,.
\]
Since the global vacuum is $SL(2,{\mathbb R})$ invariant, this state
should also be annihilated by the global Hamiltonian $\hat{R} \otimes
\identity + \identity \otimes \hat{R}$ and the Poincar\'e Hamiltonian
$\identity \otimes \hat{H}$.

What does this mean as far as constructing local bulk operators?  The
expressions we derived in section \ref{BHsmear} are perfectly
applicable.  They imply that to put a bulk operator behind the horizon
of the black hole we need boundary operators that act non-trivially on
both copies of the Hilbert space.  Related observations were made in
\cite{Maldacena:2001kr}.

Let us summarize the picture we have developed.  A bulk operator
outside the horizon corresponds to a non-local operator that acts on a
single copy of the Hilbert space.  As the bulk point approaches the
future (past) horizon the smearing function extends to cover an
infinite range of coordinate time on the boundary: it has support as
$t \rightarrow + \infty$ ($t \rightarrow - \infty$).\footnote{A
similar property held in the Poincar\'e patch: the signal of the
Poincar\'e horizon was that the smearing function extended to
Poincar\'e time $T \rightarrow \pm \infty$.}  To insert a bulk
operator behind the horizon we need a non-local operator that acts on
both copies of the Hilbert space.

What does this mean from the point of view of an observer outside the
black hole, who can only interact with a single copy of the CFT?  Such
an observer must trace over the second copy of the CFT.  If no
operators are inserted behind the horizon then the trace leads to a
precisely thermal density matrix that describes the black hole.  But
operator insertions behind the horizon act on the other copy of the
CFT, and modify the resulting density matrix.  In general these
modifications will not have a thermal character.  Thus, from the point
of view of an outside observer, {\em operator insertions behind the
horizon are seen as non-thermal modifications to the black hole
density matrix}.

\section{Higher dimensional black holes}

In the previous section we saw that in order to describe an object
inside the Rindler horizon of a two-dimensional black hole one needs a
non-local operator that acts on both copies of the Hilbert
space.\footnote{Equivalently one needs an operator that acts both at
real time $t$ and at complex time $t+i\beta/2$.}  In this section we
show that a similar property holds for a general AdS black hole.
Analogous arguments can be made for bulk points outside of the
Poincar\'e patch.

A local field $\phi(U,V,\Omega)$ anywhere in the extended Kruskal
diagram can be expanded in terms of Kruskal creation and annihilation
operators.
\begin{equation}
\phi=\sum_{i}f_{i}(U,V,\Omega) a^{i}_{K} +f^{*}_{i}(U,V,\Omega)
a^{i\dagger}_{K}
\end{equation}
The Kruskal creation and annihilation operators can be expressed in
terms of left and right creation and annihilation operators using
Bogolubov coefficients.
\begin{eqnarray}
&&a^{i}_{K} = \alpha^L_{ij}a^{j}_{L}+\beta^L_{ij}a^{j\dagger}_{L}
+\alpha^R_{ij}a^{j}_{R}+\beta^R_{ij}a^{j\dagger}_{R}\\
\nonumber
&&a^{i\dagger}_{K} = \alpha^{L}_{ij}{}^*a^{j\dagger}_{L}+\beta^{L}_{ij}{}^*a^{j}_{L}
+\alpha^{R}_{ij}{}^*a^{j\dagger}_{R}+\beta^{R}_{ij}{}^*a^{j}_{R}
\end{eqnarray}
Since the left and right creation and annihilation operators can be
written as a Fourier transform of operators in one of the copies of the
CFT,
\begin{eqnarray}
a^{j}_{L}\sim \int dt d^{d}x \, e^{-i\omega_{j}t+ik_{j}x}{\cal O}_{L}(t,x) \\
\nonumber
a^{j}_{R}\sim \int dt d^{d}x \, e^{-i\omega_{j}t+ik_{j}x}{\cal O}_{R}(t,x) 
\end{eqnarray}
we see that a local bulk field anywhere in the Kruskal diagram can be
represented as a linear combination of two operators, one acting on
the left and one acting on the right.  If the bulk point is in either
the left or the right region then its representation reduces to a
single operator on the appropriate copy.

\section{Conclusions}

In this paper we related local bulk fields in AdS to non-local
operators on the boundary.  In global coordinates we found
AdS-covariant smearing functions with support purely at spacelike
separation.  We then showed how to represent bulk operators that are
inserted behind the Poincar\'e horizon, or inside the horizon of a
black hole.  By construction these boundary operators reproduce all
bulk correlation functions; they therefore respect bulk locality.
Light-cone singularities in the bulk arise from the UV behavior of the
boundary theory.  Although we concentrated on AdS${}_2$, similar
results hold in higher dimensions \cite{ToAppear}.

It is curious that local operators in the interior of a black hole
correspond to boundary operators that act on both copies of the
thermofield double.  Note that in the thermal AdS phase, such
operators do not exist: any operator that acts on both copies of the
CFT will at best be bilocal in the (disconnected) bulk spacetime.
It is only in the black hole phase that operators which act on both
copies of the CFT can be local in the bulk.

One can even give a boundary description of the fact that the black
hole time coordinate switches from timelike to spacelike at the
horizon.  In the right Rindler wedge, or more generally whenever the
Killing vector ${\partial \over \partial t}$ is timelike, chronology
in the bulk corresponds to chronology on the boundary: bulk operators
which are inserted at the same spatial position but different values
of $t$ correspond to smeared operators on a single boundary that are
related by time translation.  This holds in the left and right Rindler
wedges, inside the Poincar\'e horizon, or even globally if one uses
global time.  But for bulk points inside the horizon, a Rindler time
translation will move the left and right boundary operators in
opposite directions.  This corresponds to the fact that ${\partial
\over \partial t}$ is spacelike inside the horizon.

There are many open questions and directions for future work.  Our
construction is applicable in the limit of semiclassical supergravity;
it would be extremely interesting to understand $1/N$ and $\alpha'$
corrections.  Non-local operators in the CFT should provide a new tool
to study bulk phenomena such as black hole singularities
\cite{Kraus:2002iv, Fidkowski:2003nf}, the interplay of holography and
locality \cite{Lowe:1999pk}, or even non-local deformations of AdS \cite{Aharony:2005sh}.
But at a more fundamental level, one might ask: purely from the
boundary CFT point of view, what if anything would allow one to
identify a particular set of non-local operators as appropriate for
describing bulk spacetime?

\bigskip
\goodbreak
\centerline{\bf Acknowledgements}
\noindent
We are grateful to Nori Iizuka and Lenny Susskind for valuable
discussions.  AH and DK are supported by DOE grant DE-FG02-92ER40699.
The research of DL is supported in part by DOE grant
DE-FE0291ER40688-Task A.  DK, DL and GL are supported in part by
US--Israel Bi-national Science Foundation grant \#2000359.

\appendix
\section{AdS isometries}

We review the relationship between isometries of AdS${}_2$ and
conformal transformations on the boundary.  For a more extensive
discussion see \cite{Gibbons:1998fa,Cadoni:1999ja}.

AdS${}_2$ can be embedded as a hypersurface
\[
-(X^0)^2 - (X^1)^2 + (X^2)^2 = -R^2
\]
in ${\mathbb R}^{2,1}$ with a $(--+)$ metric.  Isometries of AdS${}_2$
arise from Lorentz transformations of the embedding space, with
generators
\[
J_{\mu \nu} = \eta_{\mu \sigma} X^{\sigma} \partial_{\nu} - \eta_{\nu \sigma} X^{\sigma} \partial_{\mu}\,.
\]
For example the $SL(2,{\mathbb R})$-invariant distance function is
\[
\sigma(x \vert x') = {1 \over 2 R^2} (X - X')_\mu (X - X')^\nu + 1
\]
and the antipodal map is simply $A \, : \, X^\mu \rightarrow - X^\mu$.

We would like to understand the action of these isometry generators on the
boundary in different coordinate systems (global, Poincar\'e and
Rindler).  These coordinate systems are defined as follows
\cite{Balasubramanian:1998sn}:
\begin{itemize}
\item{Global: $\left\{
\begin{array}{lcl}
X^0 &=&R \sec \rho \cos \tau \\
X^1 &=&R \sec \rho \sin \tau \\ 
X^2 &=&R \tan \rho
\end{array}
\right.$
        
with metric: $ds^2 = \frac{R^2}{\cos^2 \rho} (-d\tau^2 + d\rho^2)$,}
        
\item{Poincar\'e: $\left\{
\begin{array}{lcl}
X^0 &=& (R^2 + Z^2 - T^2)/2Z \\
X^1 &=& R T / Z \\
X^2 &=& (R^2 - Z^2 + T^2)/2Z
\end{array}
\right.$
        
with metric: $ds^2 = \frac{R^2}{Z^2} (-dT^2 + dZ^2)$,}
        
\item{Rindler (with arbitrary dimensionless M): $\left\{
\begin{array}{lcl}
X^0 &=& r/\sqrt{M} \\
X^1 &=& \sqrt{\frac{r^2}{M} - R^2} \sinh(\sqrt{M}t/R) \\
X^2 &=& \sqrt{\frac{r^2}{M} - R^2} \cosh(\sqrt{M}t/R)
\end{array}
\right.$
        
with metric: $ds^2 = -\left({r^2\over R^2} - M\right) dt^2 + \left({r^2\over R^2} - M\right)^{-1} dr^2$.}

\end{itemize}
Near the boundary ($\rho \rightarrow \pi/2, z \rightarrow 0, r
\rightarrow \infty$) the isometry generators approach
\begin{itemize}
\item{Global: $\left\{
\begin{array}{lcl}
J_{01} &=& -\partial_\tau \\
J_{02} &=& \sin \tau \partial_\tau \\ 
J_{12} &=& -\cos \tau \partial_\tau
\end{array}
\right.$}
\item{Poincar\'e: $\left\{
\begin{array}{lcl}
J_{01} &=& - \frac{1}{2R} (T^2 + R^2) \partial_T \\
J_{02} &=& T \partial_T \\
J_{12} &=& \frac{1}{2R} (T^2 - R^2) \partial_T
\end{array}
\right.$}
\item{Rindler: $\left\{
\begin{array}{lcl}
J_{01} &=& - {R \over \sqrt{M}}\cosh (\sqrt{M}t/R) \partial_t \\
J_{02} &=& {R \over \sqrt{M}} \sinh (\sqrt{M}t/R) \partial_t \\
J_{12} &=& -\frac{R}{\sqrt{M}} \partial_t~.
\end{array}
\right.$}
\end{itemize}
Here we are keeping only the leading (divergent) behavior of the vector
field near the boundary.

In Poincar\'e coordinates the isometries give rise to the usual
representation for the conformal generators on a line,
\bea
\hat{H} &=& i \partial_T = - (i/R) (J_{01} + J_{12}) \\ \nonumber
\hat{D} &=& i T \partial_T = i J_{02} \\ \nonumber
\hat{K} &=& i T^2 \partial_T = -iR (J_{01} - J_{12})~.
\eea
Rather than adopting $\hat{H}$ as the Hamiltonian one can use a
different linear combination of the conformal generators to evolve in
time \cite{QM}.  This is exactly what is done when using the other two
coordinate systems:

Global:
$i \partial_\tau = \frac{1}{2} (R \hat{H} + \frac{1}{R} \hat{K}) = \hat{R}$

Rindler:
$i \partial_t =  \frac{\sqrt{M}}{2R} (R\hat{H} - \frac{1}{R}\hat{K}) =
- {\sqrt{M} \over R} \hat{S}.$

\noindent
Here $\hat{R}$ (not to be confused with the AdS radius) and $\hat{S}$
are compact rotation and non-compact boost generators, respectively.
One can show that when evolving in time using non-compact generators
(such as $\hat{S}$), one cannot cover the entire range $-\infty < T <
\infty$.  This is exactly what happens in Rindler coordinates --
Rindler time $t$ only covers half of the boundary of the Poincar\'e
patch.  When evolving with compact generators such as $\hat{R}$, one
does cover the entire range of $T$ (as global coordinates do)
\cite{QM}.

\section{AdS vacua}

At various points in the paper we made use of the equivalence between
the global, Poincar\'e and Hartle-Hawking vacua
\cite{Lifschytz:1993eb,Danielsson:1998wt,SpradlinStrominger}.  In this
appendix we review these results and show how they relate different
boundary fields.

We begin with the equivalence between the global and Poincar\'e vacua.
Bulk modes that are positive frequency with respect to global time are
also positive frequency with respect to Poincar\'e time
\cite{Danielsson:1998wt}.  This implies that the two bulk vacua are
equivalent.  To understand the corresponding relationship between the
global and Poincar\'e boundary fields, let us start with a global
boundary field that is purely positive frequency
\be
\label{GlobalPos}
\phinotgR{+}(\tau) = \sum_{n = 0}^\infty c_n e^{-i \omega_n \tau}~.
\ee
Near the boundary
\[
Z^\Delta \phinotp{} \sim \cos^\Delta \rho \, \phinotgR{}
\]
so the corresponding Poincar\'e boundary field is equal to the global field times a Jacobian
\be
\label{Jacobian}
\phinotp{}(T) = \lim_{\rho \rightarrow \pi/2} \left({\cos \rho \over Z}\right)^\Delta \phinotgR{+}(\tau) =
\left({2 R \over T^2 + R^2}\right)^\Delta \phinotgR{+}(\tau)~.
\ee
Here $\tau = 2 \tan^{-1}(T/R) = -i \log {1 + i T / R \over 1 - i T / R}$ so
\be
\phinotp{}(T) = \left({2 \over R}\right)^\Delta \sum_{n = 0}^\infty c_n \, {\!\!\!\!(1 - iT/R)^n \over (1 + iT/R)^{n + 2\Delta}}\,.
\ee
The Poincar\'e boundary field is analytic in the lower half of the
complex $T$ plane, so its Fourier transform
\[
\widetilde{\phi}_0^{\it Poincare}(\omega) = \int dT e^{i \omega T} \phinotp{}(T)
\]
vanishes if $\omega < 0$.  Thus it follows from the equivalence of the bulk Poincar\'e and global vacua that a positive-frequency global boundary
field corresponds to a positive-frequency Poincar\'e boundary field.
We used this in section \ref{behindp}  to go from (\ref{PoinGlobNonZ}) to
(\ref{PoinPoinNonZ}).  Note that the Jacobian appearing in
(\ref{Jacobian}) is crucial: the global boundary field
(\ref{GlobalPos}) by itself, when expressed in terms of Poincar\'e
time, is {\em not} positive-frequency.

Now let us turn to the equivalence between the global and
Hartle-Hawking vacua.  For these purposes it is convenient to introduce
null Kruskal coordinates
\be
u = \tan {\tau + \rho \over 2} \qquad v = \tan {\tau - \rho \over 2}
\ee
in which
\be
ds^2 = - {4 R^2 du dv \over (1 + uv)^2}\,.
\ee
These coordinates cover the region shown in Fig.~4; the restriction
$-\pi/2 < \rho < \pi/2$ corresponds to $uv > -1$.  For $u < 0$ note
that points with $uv = -1$ make up the boundary of the left Rindler
wedge, while for $u > 0$ points with $uv = -1$ make up the boundary of
the right Rindler wedge.

\begin{figure}
\centerline{\epsfig{file=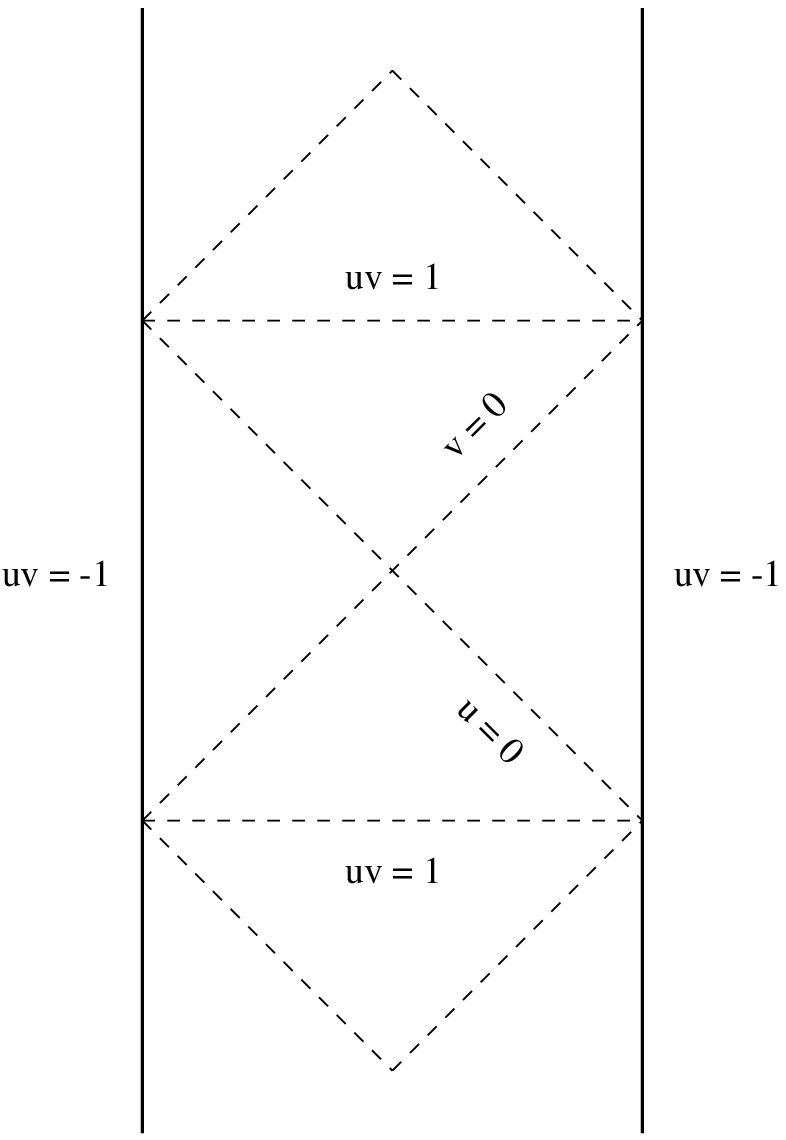} \epsfig{file=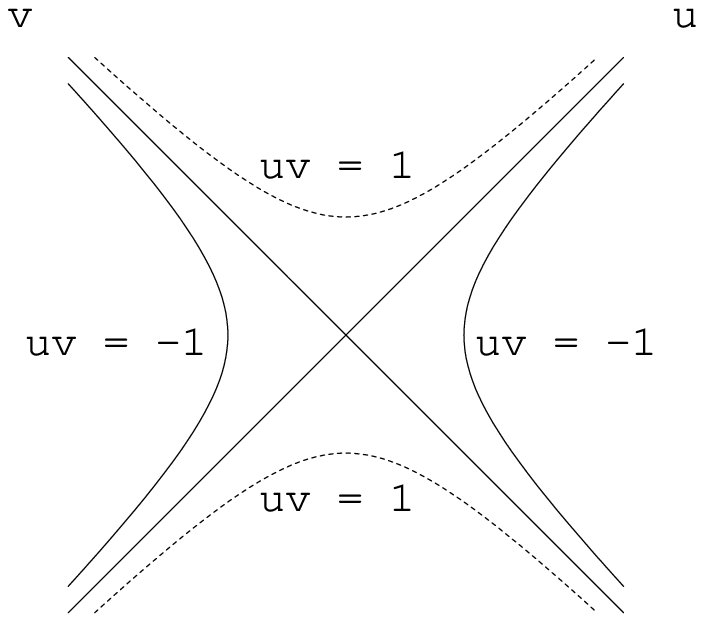}}
\caption{The Kruskal patch, indicated on the AdS Penrose diagram
(left) and in the $uv$ plane (right).}
\end{figure}

Working in the right Rindler wedge ($u > 0$ and $v < 0$) mode
solutions with normalizable fall-off near the right boundary of AdS
are
\be
\label{KruskalMode}
\phi(u,v) = u^{-i\omega} (1 + uv)^\Delta F(\Delta, \Delta - i \omega, 2\Delta, 1+uv)\,.
\ee
Here $-\infty < \omega < \infty$ is a parameter which can be
understood as the frequency measured with respect to Rindler time,
since under the isometry corresponding to a Rindler time translation
\[
u \rightarrow e^\lambda u \qquad v \rightarrow e^{-\lambda} v
\]
these modes transform by a definite phase $\phi \rightarrow e^{-i
\omega \lambda} \phi$.  Note that $\omega$ does not correspond to the
frequency measured with respect to a Kruskal time coordinate.

We want to find a set of positive-frequency Kruskal
modes\footnote{Positive frequency only in the sense that they multiply
annihilation operators in the mode expansion of the field.  They do
not transform by a definite phase under a Kruskal time translation.}
for which (by construction) the Kruskal vacuum will be equivalent to
the global vacuum.  This depends on making the correct analytic
continuation from right to left.  To do this we first write the modes
(\ref{KruskalMode}) in a more symmetric form, using a transformation
formula for the hypergeometric function
\bea
\nonumber
\phi(u,v) & = & (1 + uv)^\Delta \biggl[
{\Gamma(2\Delta) \Gamma(i\omega) \over \Gamma(\Delta) \Gamma(\Delta + i\omega)} u^{-i\omega}
F(\Delta, \Delta - i \omega, 1 - i\omega,-uv) + \\
\label{KruskalModes2}
& & \qquad \qquad {\Gamma(2\Delta) \Gamma(-i\omega) \over \Gamma(\Delta) \Gamma(\Delta - i\omega)} (-v)^{i\omega}
F(\Delta, \Delta + i \omega, 1 + i\omega,-uv)\biggr]~.
\eea
This makes it clear that, aside from the prescribed behavior near the
boundary of AdS, the modes have branch points on the horizon (at $u =
0$ or $v = 0$).  A positive frequency Kruskal mode is defined by
analytically continuing from right to left going through the lower
half of the complex $u$ and $v$ planes, while a negative frequency
Kruskal mode is defined by analytically continuing through the upper
half $u$ and $v$ planes.\footnote{This is the same prescription used
in Rindler space \cite{BirrellDavies} or, more generally, on a
bifurcate Killing horizon \cite{Israel:1976ur}.}

It is straightforward to obtain the corresponding analyticity
properties working in terms of boundary fields.  In Kruskal
coordinates we define the boundary field by
\be
\phi_0^{\it Kruskal}(u) = \lim_{uv \rightarrow -1} {\phi(u,v) \over (1 + uv)^\Delta}~.
\ee
The relation between Kruskal and global boundary fields is then
\be
\phi_0^{Kruskal}(u) =\left\lbrace
\begin{array}{ll}
\lim\limits_{uv\to -1} \frac{\cos^\Delta \rho}{(1+uv)^\Delta} \phi_0^{\it global,~R}(\tau) & \hbox{\rm for $u > 0$} \\
\lim\limits_{uv\to -1} \frac{\cos^\Delta \rho}{(1+uv)^\Delta} \phi_0^{\it global,~L}(\tau) & \hbox{\rm for $u < 0$}~.
\end{array}
\right.
\label{kgrel}
\ee
A positive frequency global boundary field takes the form
\be
\label{PosGlobBdy}
\phinotgR{+}(\tau)=\sum_{n=0}^\infty c_n e^{-i \omega_n \tau}~,\qquad
\phinotgL{+}(\tau)=\sum_{n=0}^\infty c_n (-1)^n e^{-i \omega_n \tau}~.
\ee
Note that
\be
\label{GlobPosCond}
\lim_{\tau \rightarrow +\pi/2} \phinotgL{+}(\tau) = e^{-i \pi \Delta} \lim_{\tau \to -\pi/2} \phinotgR{+}(\tau)~.
\ee
Inserting (\ref{PosGlobBdy}) into (\ref{kgrel}) we find that a
positive frequency global boundary field maps to a Kruskal boundary
field given by
\be
\phi_0^{\it Kruskal}(u) = \sum_{n=0}^\infty c_n   \frac{(-1)^{n/2}(-iu)^\Delta (u+i)^n}{(u-i)^{2\Delta+n}}~.
\ee
In this expression the branch cut beginning at $u = 0$ is taken to lie
in the upper half complex $u$ plane in order to reproduce
(\ref{GlobPosCond}).  Thus a positive frequency global mode
corresponds to a Kruskal mode that is analytic and bounded in the
lower half of the complex $u$ plane.  We take this analyticity
condition to define a positive frequency Kruskal boundary mode; this
makes the global and Kruskal vacua equivalent on the boundary (just as
they were equivalent in the bulk).

With this definition, from (\ref{KruskalModes2}) a positive-frequency
Kruskal mode can be characterized by the boundary behavior
\[
\phinotk{+}(u) = \left\lbrace
\begin{array}{ll}
u^{-i\omega} & \hbox{\rm for $u > 0$} \\
e^{-\pi\omega} (-u)^{-i\omega} & \hbox{\rm for $u < 0$}
\end{array}
\right.
\]
while a negative-frequency Kruskal mode is characterized by
\[
\phinotk{-}(u) = \left\lbrace
\begin{array}{ll}
u^{-i\omega} & \hbox{\rm for $u > 0$} \\
e^{+\pi\omega} (-u)^{-i\omega} & \hbox{\rm for $u < 0$}\,.
\end{array}
\right.
\]
At this point it is convenient to switch to Rindler coordinates.  The relation is
\[
u = \pm \left({r - \sqrt{M} R \over r + \sqrt{M} R}\right)^{1/2} e^{\sqrt{M} t/R} \qquad
v = \mp \left({r - \sqrt{M} R \over r + \sqrt{M} R}\right)^{1/2} e^{-\sqrt{M} t/R} \qquad
\]
where the upper (lower) sign holds in the right (left) Rindler wedge.
The Jacobian for switching from Kruskal to Rindler is a constant,
\[
\phinotr{} = \lim_{r \rightarrow \infty} (r(1 + uv))^\Delta \phinotk{} = (2 \sqrt{M} R)^\Delta \phinotk{}\,,
\]
so we can characterize Rindler modes which are positive or negative
frequency with respect to Kruskal time as having the behavior
\[
\phinotrR{\pm} \sim e^{-i \omega t} \qquad
\phinotrL{\pm} \sim e^{\mp \beta \omega / 2} e^{-i \omega t}
\]
where $\beta = 2 \pi R / \sqrt{M}$ is the inverse temperature of the
black hole.  Finally, this lets us define a projection operator which
picks out the part of the Rindler boundary field which is positive or
negative frequency with respect to Kruskal time.  Let
\[
\widetilde{\phi}_0^{\it Rindler,\,L/R}(\omega) = \int_{-\infty}^\infty dt \, e^{i \omega t} \phi_0^{\it Rindler,\,L/R}(t)
\]
be the Fourier transform of the Rindler boundary field.  Then Rindler
boundary fields which are positive or negative frequency with respect
to Kruskal time are given by
\bea
\label{RindlerKruskalPos}
&& \phinotrR{\pm}(t) = \int_{-\infty}^\infty {d\omega \over 2\pi} f_\pm(\omega) e^{-i \omega t} \\
\nonumber
&& \phinotrL{\pm}(t) = \int_{-\infty}^\infty {d\omega \over 2\pi} f_\pm(\omega) e^{\mp \beta \omega / 2} e^{-i \omega t}
\eea
where the conditions
\bea
&& f_+ + f_- = \widetilde{\phi}_0^{\it Rindler,\,R} \\
\nonumber
&& f_+ e^{- \beta \omega / 2} + f_- e^{\beta \omega / 2} = \widetilde{\phi}_0^{\it Rindler,\,L}
\eea
fix
\bea
&& f_+(\omega) = {\widetilde{\phi}_0^{\it Rindler,\,R}(\omega) e^{\beta \omega /2} - \widetilde{\phi}_0^{\it Rindler,\,L}(\omega)
\over 2 \sinh \beta \omega / 2} \\
\nonumber
&& f_-(\omega) = {\widetilde{\phi}_0^{\it Rindler,\,L}(\omega) - \widetilde{\phi}_0^{\it Rindler,\,R}(\omega) e^{-\beta \omega / 2}
\over 2 \sinh \beta \omega / 2}~.
\eea
Multiplying by the Rindler-to-global Jacobian
\be
\lim_{r \rightarrow \infty} {1 \over (r \cos \rho)^\Delta} = \left({\cosh(\sqrt{M} t / R) \over \sqrt{M} R}\right)^\Delta
\ee
the boundary fields (\ref{RindlerKruskalPos}) become positive or
negative frequency with respect to global rather than Kruskal time.
They can then be substituted in (\ref{GeneralRind}) to obtain the
Rindler smearing function for a point inside the horizon.


\end{document}